\documentclass[%
reprint,
superscriptaddress,
 amsmath,amssymb,
 aps,
prl,
longbibliography
]{revtex4-1}

\usepackage{xcolor}

\usepackage{braket}
\usepackage{graphicx}% Include figure files
\usepackage{dcolumn}% Align table columns on decimal point
\usepackage{bm}% bold math
\begin{document}

\title{Molecular spin-probe sensing of H-mediated changes in Co nanomagnets}

\author{A. F\'{e}tida}
\author{O. Bengone}
\author{C. Goyhenex}
\author{F. Scheurer}
\affiliation{Universit\'{e} de Strasbourg, CNRS, IPCMS, UMR 7504, F-67000 Strasbourg, France\looseness=-1}
\author{R. Robles}
\affiliation{Centro de Física de Materiales CFM/MPC (CSIC-UPV/EHU), Paseo Manuel de Lardizabal 5, 20018 Donostia-San Sebasti\'an, Spain\looseness=-1}
\author{N. Lorente}
\affiliation{Centro de Física de Materiales CFM/MPC (CSIC-UPV/EHU), Paseo Manuel de Lardizabal 5, 20018 Donostia-San Sebasti\'an, Spain\looseness=-1}
\affiliation{Donostia International Physics Center (DIPC), 20018 Donostia-San Sebasti\'{a}n, Spain\looseness=-1}
\author{L. Limot}
\email{limot@ipcms.unistra.fr}
\affiliation{Universit\'{e} de Strasbourg, CNRS, IPCMS, UMR 7504, F-67000 Strasbourg, France\looseness=-1}

\begin{abstract}
The influence of hydrogen on magnetization is of significant interest to spintronics. Understanding and controlling this phenomenon at the atomic scale, particularly in nanoscale systems, is crucial. In this study, we utilized scanning tunneling microscopy (STM) combined with a nickelocene molecule to sense the spin of a hydrogen-loaded nanoscale Co island grown on Cu(111). Magnetic exchange maps obtained from the molecular tip revealed the presence of a hydrogen superstructure and a 90$^{\circ}$ rotation of the magnetization compared to the pristine island. \textit{Ab initio} calculations corroborate these observations, indicating that hydrogen hybridization with Co atoms on the island surface drives the spin reorientation of the island. This reorientation is further reinforced by hydrogen penetration into the island that locates at the Co/Cu interface. However, the subsurface sensitivity of the magnetic exchange maps indicate that this effect is limited. Our study provides valuable microscopic insights into the chemical control of magnetism at the nanoscale.
\end{abstract}
\date{\today}


 \maketitle

\noindent\textbf{INTRODUCTION}\\
Magnetic anisotropy determines the orientation of spins in metallic thin films and multilayers, thereby constituting a critical aspect of spintronic device functionality. Among various approaches for controlling magnetic anisotropy, the incorporation of mobile ionic species in thin films through voltage-gated transport stands out as promising. Hydrogen, in particular, has proven superior to other ionic species, offering non-destructive and rapid toggling of magnetic anisotropy of multilayered heterostructures \cite{Tan2019,Lee2020,Kossak2023}. The rationale for hydrogen loading is built upon the observation that even subtle interactions, such as charge transfer between hydrogen and a metal atom, can trigger changes in the magnetic anisotropy and effective magnetic moment of the atom \cite{Dubout2015,Khajetoorians2015,Jacobson2015,Gonzalez2016,Jacobson2017,Steinbrecher2021}, while also affecting the exchange coupling among atoms \cite{Hjorvarsson1997,Leiner2003}. The magnetization orientation in ultrathin films can be modified by the adsorption and penetration of hydrogen \cite{Hjorvarsson1997,Leiner2003,Sander2004,Munbodh2011,Santos2012}. This effect is observed not only in collinear magnetic films, but also in noncollinear ones \cite{Chen2021,Wang2022}, where hydrogen has the ability to stabilize skyrmion states \cite{Hsu2018,Chen2022}. The position and concentration of hydrogen in and on the host metal are important factors in this process \cite{Hsu2018,Klyukin2020}. While recent advancements have enabled microscopic-level imaging of hydrogen \cite{Graaf2020}, these measurements have not yet been associated with magnetism.

Single hydrogen molecules are commonly studied using STM. Hydrogen molecules assemble into coverage-dependent lattices on metal surfaces \cite{Natterer2013a} and have a characteristic vibrational structure facilitating their identification \cite{Natterer2013a,Li2013}. They can also be manipulated by the STM tip \cite{Lotze2012,Merino2019,Wang2022b}, adding chemical functionality to the microscope \cite{Weiss2010}. Dissociative adsorption of hydrogen molecules has also been reported on magnetic surfaces, where hydrogen atoms then form superlattices \cite{Sicot2008a,Park2017}. However, conducting spin-sensitive STM measurements on hydrogen-exposed magnets remains a challenging endeavor \cite{Park2017,Hsu2018}, as it necessitates a magnetic tip, which, similar to the surface, is susceptible to hydrogen contamination \cite{Hofer2008}. Rigorous monitoring of the tip apex status is essential, preferably with an external magnetic field.\\

\begin{figure}
 \includegraphics[width=1\columnwidth]{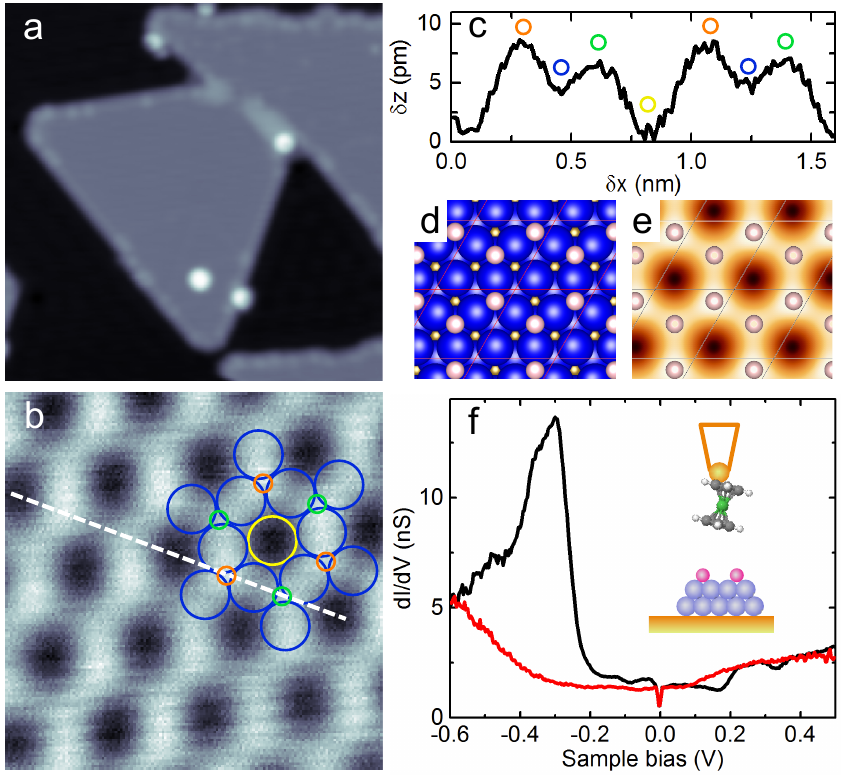}
  \caption{(a) Constant-current image of H-covered Co bilayer islands on Cu(111) (image size: $20\times20$~nm$^2$, sample bias: $50$~mV, tunnel current: $50$~pA). The bright spots are Nc molecules. (b) Close-up view of the island ($1.7\times1.7$~nm$^2$, sample bias: $50$~mV, tunnel current: $50$~pA) with $2$H-$(2\times2)$ superstructure [Co(I): yellow, Co(II): blue, H(I): green, H(II): orange]. The Co network corresponds to that of the pristine island before hydrogen exposure. (c) Height profile acquired along the dashed line indicated in (b). Coloured circles indicate the position of the H and Co atoms. (d) Model structure of $2$H-$(2\times2)$ (Co: blue, H: white) and (e) corresponding DFT-computed constant-current topography (simulation done at $100$ mV). Circles represent H atoms. (f) Typical $dI/dV$ spectra acquired on H-contaminated (solid red line) and pristine islands (solid black line). Feedback loop opened at $1.2$~nA and $0.5$~V. Inset: sketch of the tunnel junction.}
\label{F1}
\end{figure}

\noindent\textbf{RESULTS AND DISCUSSION}\\
To overcome this challenge, we passivated the apex of a copper-coated W tip with a magnetic nickelocene molecule [Ni(C$_5$H$_5$)$_2$, denoted as Nc, see Fig.~\ref{F1}f], and focused on a model system comprising a nanoscale Co magnet that was exposed to hydrogen. Owing to the direct exchange coupling between the Nc-terminated tip and the Co surface of the magnet, we employed the inelastic tunnel current to monitor the orientation of the surface magnetization with atomic-scale sensitivity \cite{Verlhac2019,Fetida2024}. Magnetic exchange maps obtained with the molecular tip reveal that hydrogen is predominantly located on the magnet's surface, forming a hydrogen superstructure, while the subsurface sensitivity of the magnetic probe tip indicates weaker hydrogen penetration into the magnet. This results in a 90$^{\circ}$ rotation of the magnetization compared to the hydrogen-free state. These observations, corroborated by \textit{ab initio} calculations, represent an advancement in our comprehension of how hydrogen alters the magnetism of a nanoscale system at the atomic level.

The sample consisted of Co islands grown on Cu(111), which are two atomic layers high and exhibit a triangular-like structure (Fig.\ref{F1}a) \cite{Diekhoner2003,Pietzsch2004,Rastei2007}. The apparent height is $325\pm 10$ pm, and typical lateral sizes range from $10$ to $30$ nm. After island growth, Nc is deposited onto the sample, which is then exposed to hydrogen (see Materials and Methods for details). To assess H-adsorption onto Co islands, a Nc-tip is prepared by transferring a single Nc from the surface to a mono-atomically sharp tip. Imaging can then be conducted via two operating modes. The first mode involves tunneling with parameters $50$ mV and $50$ pA, for which the exchange interaction with the Co surface is absent (Fig.~S1, for details, see the Supporting Materials)~\cite{SM}. In this mode, which we define as a high-bias mode, the Nc-tip functions as a standard STM tip. The second mode, which we define as a low-bias mode, involves tunneling with parameters $1$ mV and $100$ pA, where the Nc-tip is $50$ pm away from the magnetic surface for the exchange interaction to be active. This mode carries magnetic information. Switching between high- and low-bias allows us to visualize the electronic and magnetic properties, respectively.\\ 

\noindent\textbf{Hydrogen superstructure on a Co bilayer island}\\
To determine the hydrogen coverage on a nanoscale island, we employed the Nc-tip in the high-bias mode. Figure~\ref{F1}a shows a typical image of an island after hydrogen exposure (see Materials and Methods), which appears similar to a pristine island. To visualize the presence of hydrogen atoms, it is necessary to zoom in on the island. A close-up view of a Co island is presented in Fig.\ref{F1}b, along with an overlaid simulation of the Co network of the pristine island. Hydrogen atoms are positioned on the hollow sites of the Co surface and alternate between two distinct hollow sites, resulting in a $2$ pm difference in their apparent height (Fig.\ref{F1}c). The hydrogen coverage is 0.5 monolayers (ML), suggesting a hydrogen arrangement in a $2$H-$(2\times2)$ superstructure (Fig.~\ref{F1}d). The unit cell of the superstructure contains two Co atoms: one at the center of a hexagon without neighboring H atoms [noted Co(I), yellow in Fig.~\ref{F1}b], and a second, which has two neighboring H atoms [noted Co(II), blue]. The apparent height of Co(II) is $4$ pm greater than that of Co(I). To corroborate these observations, DFT calculations were carried out, yielding good agreement with the experimental images (Fig.~\ref{F1}e). The strongest corrugation in the computed image corresponds to hydrogen in an \textit{hcp} site [noted H(I), green], while the weaker corrugation corresponds to hydrogen in an \textit{fcc} site [noted H(II), orange].

This assignment, however, needs to be validated also by tunneling spectroscopy. The $1$H-$(2\times2)$ superstructure, where one H atom is present in a $(2\times2)$ Co unit cell corresponding to a $0.25$ ML coverage, is in fact indistinguishable from its $2$H counterpart in the computed images (Fig.~S2). Figure~\ref{F1}f presents a typical $dI/dV$ spectrum of a pristine Co island, revealing a prominent peak at $-0.33$ eV, arising from minority $d_{z^2}$ states hybridized with $s-p$ states \cite{Fetida2024}. The sharp feature at zero-bias is due to the inelastic tunnel current of the Nc-tip. After H exposure, the peak shifts out of our energy window, \textit{i.e.}, below $-0.6$ eV. This behavior is consistent with the computed LDOS for both Co(I) and Co(II) in the $2$H-$(2\times2)$ superstructure (Fig.~S2). It contrasts with the LDOS of a $1$H-$(2\times2)$ superstructure, which maintains instead a spin-polarized $d$-structure comparable to the pristine surface (Fig.~S2). Hence, the islands in this study exhibit a $2$H-$(2\times2)$ superstructure after hydrogen exposure.\\

\begin{figure}
 \includegraphics[width=1\columnwidth]{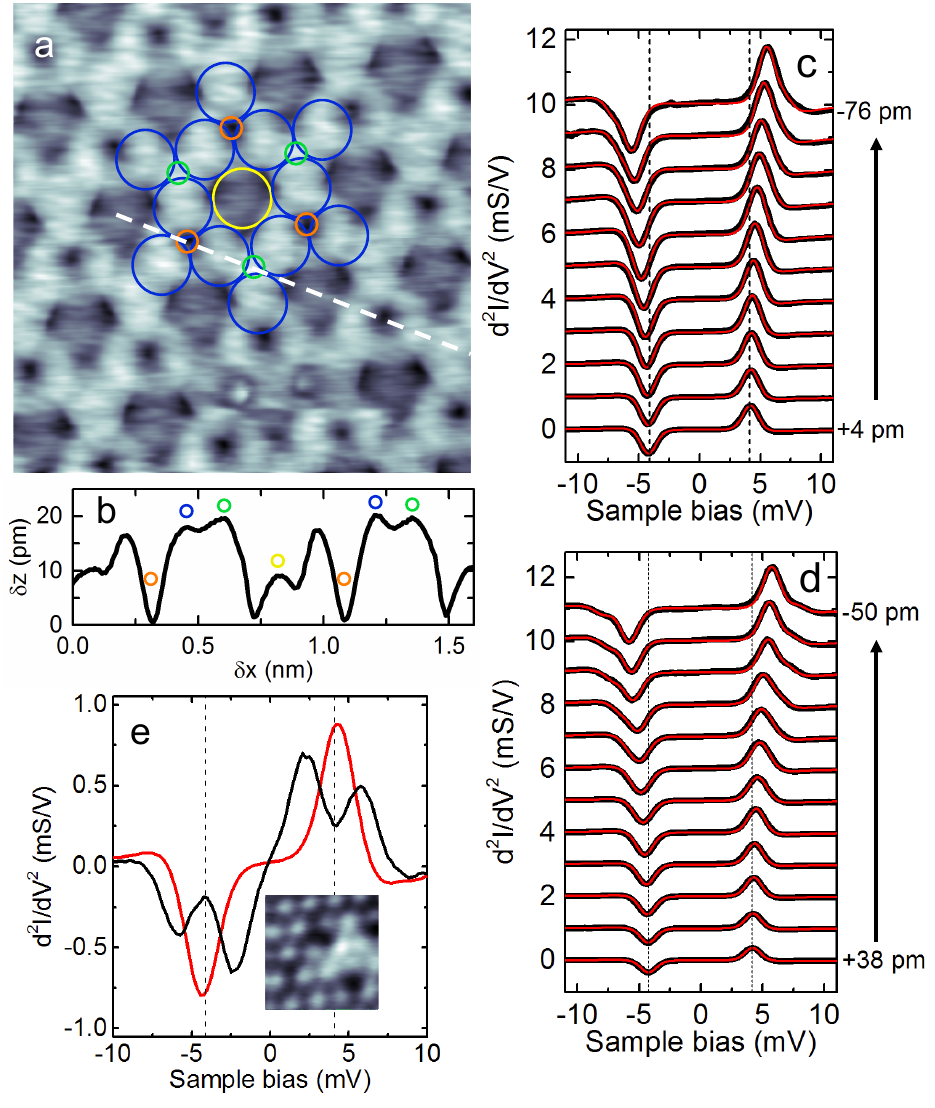}
  \caption{(a) Low-bias constant-current image of an H-covered Co bilayer island ($1.7\times1.7$~nm$^2$, $1$~mV, $100$~pA) with simulated $2$H-$(2\times2)$ superstructure [Co(I): yellow, Co(II): blue, H(I): green, H(II): orange]. (b) Height profile acquired along the dashed line indicated in (a). Coloured circles indicate the position of the H and Co atoms. Panels (c) and (d) present the $d^2I/dV^2$ spectra above Co(I) and Co(II) atoms for different tip-Co distances. Feedback loop opened above Co(I). Red lines: Simulations based on a dynamical scattering model \cite{Ternes2015}. (e) $d^2I/dV^2$ spectra acquired before and after hydrogen removal (solid red line: before, solid black line: after). The dashed lines indicate the non-shifted peak (dip) position at large distance. Inset: Image acquired after H-removal ($1\times1$~nm$^2$, $1$~mV, $100$~pA). Hydrogen removal is achieved through several $1$ V-pulses applied for a few seconds at a tunneling current of $10$ $\mu$A. It specifically targets the island beneath the tip, without perturbing neighboring islands. After the procedure, the Nc-tip is lost and must be re-prepared.}  \label{F2}
\end{figure}

\noindent\textbf{Magnetization in the presence of hydrogen}\\
To investigate the magnetism of the $2$H-$(2\times2)$ superstructure, the surface is imaged by using the low-bias mode of the Nc-tip. The low-bias image of Fig.~\ref{F2}a differs from the one previously recorded at higher bias. The Co(I) and Co(II) atoms show a stronger contrast due to their distinct corrugation (Fig.~\ref{F2}b), with a measured difference of $8$ pm between them. The two hydrogen atoms are also visible, but their contrast is reversed compared to the high-bias image, H(I) now exhibiting a height  $20$ pm larger than H(II). To determine the orientation of the island magnetization, we recorded a series of $d^2I/dV^2$ spectra by approaching the tip above each Co atom. The zero tip-displacement ($\delta z=0$) is set to a conductance of $G=0.01 G_0$ above a Co(I) atom, where $G_0=2e^2/h$ is the quantum of conductance. Conductance versus tip-displacement traces are presented in Fig.~S3. As shown in Fig.~\ref{F2}c for Co(I) and in Fig.~\ref{F2}d for Co(II), at a tip displacement of $\delta z=0$ the spectra exhibit a dip and a peak at biases of $-4.0$ and $+4.0$ mV. These features are usual for Nc-tips above non-magnetic copper \cite{Ormaza2017a}, and indicate the absence of exchange interaction. As the tip approaches the surface, the peak (dip) for both Co atoms shifts upward (downward) in energy, indicating exchange coupling between the Nc-tip and Co. This is also visible in the 2D intensity plots acquired above both cobalt atoms (Fig.~S4). The spectra can only be reproduced using a dynamical scattering model with the magnetization of the islands laying in-plane (solid red lines in Figs.~\ref{F2}c and \ref{F2}d)~\cite{Ternes2015}. Peaks and dips at opposite voltage polarities exhibit different amplitudes (refer to Fig.~\ref{F2}e, solid red line). This disparity arises from the selection rule for spin excitation and their intensity ratio provides insight into the magnitude and sign of the spin polarization of the tunnel junction \cite{Loth2010a}. The spin polarization (P) found from our simulated spectra is constant with tip displacement with $P=-0.12$ for both Co(I) and Co(II) (Fig.~S5). The spectra differ from those of the pristine island. This becomes evident upon removing hydrogen by locally heating the island with the STM tip~\cite{Sicot2008a}. Following hydrogen removal, the Co network is restored (Fig.~\ref{F2}e), and Co atoms display a characteristic spin-split $d^2I/dV^2$ spectrum, indicating out-of-plane magnetization and nearly-zero spin polarization as in a pristine Co bilayer island ~\cite{Verlhac2019,Fetida2024}. The $2$H-$(2\times2)$ superstructure can therefore be associated to a rotation of the island magnetization, from out-of-plane to in-plane. 

\begin{figure}
 \includegraphics[width=1\columnwidth]{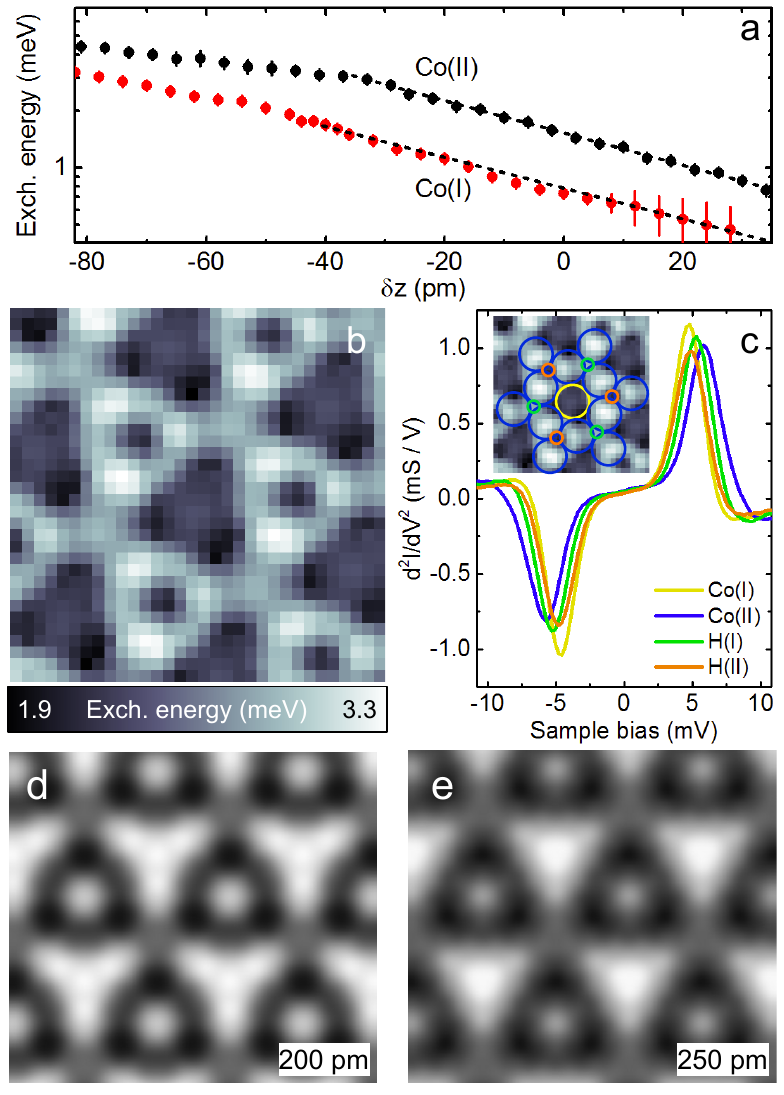}
  \caption{(a) Exchange energy versus $z$ extracted from the $d^2I/dV^2$ spectra of Co(I) and Co(II). The dashed lines represent the exponential fit described in the text. (b) Spatial map of the exchange energy acquired at a fixed distance above a Co island ($\delta z=-40$ pm). Lateral (vertical) tip drift between each spectrum acquisition was kept below 10 pm (5 pm) by dynamically tracking a local maximum. (c) $d^2I/dV^2$ spectra above Co(I), Co(II), H(I) and H(II) [Co(I): yellow, Co(II): blue, H(I): green, H(II): orange]. Computed spin density map at a fixed distance of (d) $200$ pm and (e) $250$ pm above an H atom. } 
\label{F3}
\end{figure}

%This 90$^{\circ}$ rotation of the magnetization induced by the H-exposure can be readily observed here owing to the exchange coupling of the Nc-tip to the surface. However, it is not resolved in spin-polarized STM measurements, which suffer from the weakened spin-polarized $d$-structure of the island LDOS~\cite{Park2017} and possible tip-contamination by hydrogen.

Figure~\ref{F3}a presents the distance-dependent behavior of the exchange energy derived from the spectra above Co(I) and Co(II). For $\delta z>-40$ pm, the exchange energy exhibits an exponential variation $\exp{(-\delta z/\lambda)}$ with a decay length of $\lambda=46\pm10$ pm for both atoms, similar to a Co atom of the pristine island \cite{Fetida2024}. Notably, the exchange energy above Co(II) is nearly twice as strong as the exchange energy measured above Co(I). When the tip is approached further ($\delta z<-40$ pm), the exchange energy becomes less sensitive to tip displacement, showing a smaller decay length. We assign this behavior to an Nc-hydrogen repulsion that is more pronounced on Co(II) than on Co(I) (Fig.~S3). The exchange coupling also varies across the surface. To visualize these variations, we acquire a voxel image consisting of a dataset in the form of $d^2I/dV^2 (x,y,V)$, while maintaining a fixed tip-sample distance. We then determine the exchange energy at each lateral tip position by fitting the spectra. Figure~\ref{F3}b shows the resulting magnetic exchange map, while Fig.~\ref{F3}c displays $d^2I/dV^2$ spectra at the Co and H atoms. The exchange map is similar to the low-bias image of Fig.~\ref{F2}a since the corrugation in both images is governed by the shift of the inelastic threshold energy. If the low-bias image allows quickly assessing the lateral dependence of the exchange coupling, it is done at the expense of a quantitative estimate of the exchange energy. The exchange map reveals a corrugation of the exchange energy over a $1.5$ meV range, with clear differences among Co and H atoms. However, these differences are not reflected in their magnetic moments. The DFT computed magnetic moments are $1.9$ $\mu_\text{B}$ and $1.6$ $\mu_\text{B}$ for Co(I) and Co(II), respectively, while the hydrogen atoms exhibit weak magnetism, with H(I) at $-0.02$ $\mu_\text{B}$ and H(II) at $-0.01$ $\mu_\text{B}$.

\begin{figure}
 \includegraphics[width=1\columnwidth]{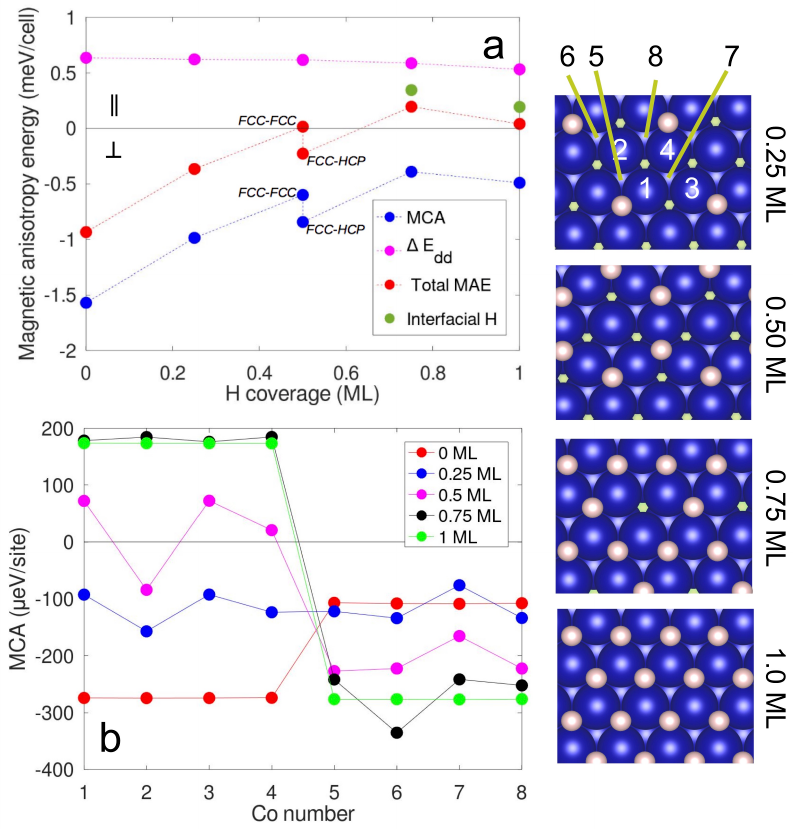}
  \caption{(a) MAE versus surface hydrogen coverage (ML). Blue: MCA Energy; magenta: Shape anisotropy, $\Delta E_{dd}$; red: Total MAE. The two additional values represented by the green circles correspond to the MAE of the configurations with $0.5$ ML of H at the surface plus $0.25$ ML and $0.50$ ML of H at the Co/Cu interface. These values correspond, respectively, to coverages of $0.75$ ML and $1$ ML. For convenience, the energy values are given in meV/cell where the considered Co $2\times 2$-cell contains $8$ Co atoms. (b) Local site analysis of the MCA. The abscissa indicate the number of the Co atom in the cell (right panels): $1-4$ for the surface layer and $5-8$ for the first Co layer.}
\label{F4}
\end{figure}

Consistent with findings on pristine Co surfaces \cite{Fetida2024}, the magnetic exchange map is reproduced by the DFT-computed spin density. Figures~\ref{F3}d-e present the spin density computations at two distinct distances from the surface, based on a $2$H-$(2\times2)$ superstructure on a Co bilayer. Remarkably, this agreement extends to H-covered Co islands of varying Co island thicknesses (Fig.~S6). The maps indicate that Co(II) exhibits a stronger spin density compared to Co(I) at all distances from the surface, aligning with the experimentally observed stronger exchange energy. The observed differences in the Co atoms, also visible in their computed LDOS (Fig.~S2), are attributed to varying hybridization with the hydrogen atoms. Additionally, the spin density of H(II) is stronger than that of H(I), confirming the contrast reversal seen experimentally for these hydrogen atoms compared to the high-bias images.\\

\noindent\textbf{Computed magnetic anisotropy energy}\\
To gain deeper insight into our findings, we have performed a computational study of hydrogen adsorption and its impact on the island's magnetization orientation. The key quantity to calculate is the Magnetic Anisotropy Energy (MAE) which is the sum of two contributions. The first contribution, known as the magnetocrystalline anisotropy Energy (MCA), stems from the spin-orbit coupling and dictates the preferred alignment direction of magnetic moments within the material arising from the crystalline structure and symmetry. The second component, termed shape anisotropy, is driven by magnetostatic dipole-dipole interactions and manifests as a tendency for the magnetic moments to align in a specific direction dictated by the object's shape. Here, a negative MAE corresponds to out-of-plane magnetization (denoted as $\perp$), while a positive value indicates in-plane magnetization (denoted as $\parallel$). Calculation details are given in Materials and Methods. To keep the calculations within practical time limits, we use the approximation of infinite bilayers of Co on a Cu(111) substrate. This approach is reasonable given that the triangular islands seen in experiments are larger than 10 nm on each side. The shape anisotropy in a triangle is then almost the same than the one of an infinite film of the same height. 

In the first step of this theoretical study, we consider only the hydrogen adsorption on top of the Co bilayer. We vary the hydrogen coverage from $0$ to $1$ ML and calculate the MAE for each coverage. Since these calculations are conducted using a $2\times2-(111)$ cell configuration, these coverages correspond to the addition of $0$ to $4$ hydrogen atoms per unit cell on the (111) surface. The H atoms are positioned in hollow \textit{fcc} sites.  For the $0.5$ ML coverage we also consider a superstructure alternating \textit{fcc} and \textit{hcp} positions as more stable, in agreement with STM experiments. The results are presented in Fig.~\ref{F4}a. For pristine Co ($0$ ML of H), the MCA is negative and exceeds the shape anisotropy, resulting in a out-of-plane magnetization ($\text{MAE}<0$). The main effect of adding hydrogen is to bring the MAE close to zero. This is driven by the MCA changing from $-1.5$ to $-0.5$ meV/cell, while the shape anisotropy remains constant at $0.6$ meV/cell. At a hydrogen coverage of $0.5$ ML, the two contributions compensate, resulting in an MAE that is nearly zero. Above $0.5$ ML, the further growth of the MCA is slower but sufficient to favor in-plane magnetization. Thus, the MCA evolution with hydrogen coverage is the driving force behind the rotation of the magnetization toward the in-plane configuration. 

A deeper understanding of the physical origin can be gained through a Co site-resolved analysis, which involves plotting the MCA for each Co site (Fig.~\ref{F4}b). As the H coverage on the surface increases, the MCA increases at the Co sites on the surface (numbered $1$ to $4$ in Fig.~\ref{F4}d). The MCA becomes large and positive above 0.5 ML, favoring an in-plane orientation of the magnetization. In contrast, the Co sites of the first layer (noted $5$ to $8$ in Fig.~\ref{F4}b), which have no hydrogen atoms as nearest neighbors, exhibit opposite behavior when the H coverage is $>0.5$ ML, thus favoring an out-of-plane magnetization. Adding up all the MCA per site yields a negative value, as shown in Fig.~\ref{F4}a. The different MCA for surface and first-layer Co atoms is attributed to their hybridization with H atoms that changes the relative importance of different $d$ orbitals of Co (Fig.~S7), as also evidenced in Pd/Co/Pd thin films \cite{Klyukin2020}.\\ 

\begin{figure}
 \includegraphics[width=1\columnwidth]{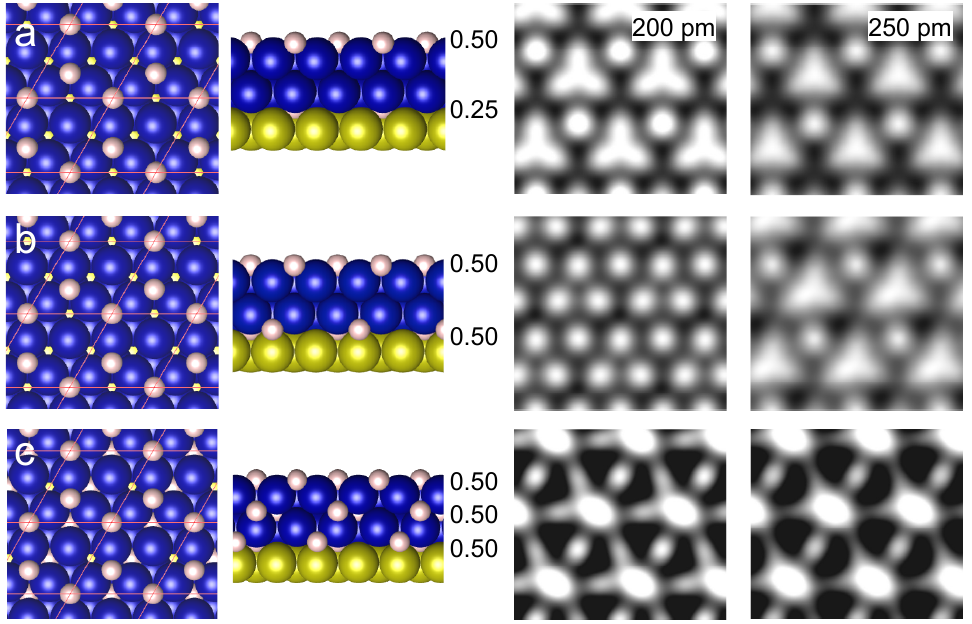}
  \caption{Top and side view of hydrogen loading in a Co bilayer possessing a $2$H-$(2\times2)$ superstructure on its surface. Two spin density maps are presented, that are computed at two fixed distances above an H atom (left: $200$, right: $250$ pm). In panel (a) 0.25 ML of hydrogen was added to the Co/Cu interface, (b) 0.50 ML of hydrogen was added to the Co/Cu interface, (c) 0.50 ML of hydrogen was added to both the first Co layer and the Co/Cu interface. The H-coverage of (a) and (b) correspond to the green circles in Fig.~\ref{F4}c.}
\label{F5}
\end{figure}

\noindent\textbf{Determining hydrogen loading in the experiment}\\
The above considerations regarding the MCA per Co site lead us to consider the possibility of H insertion as an enhancing factor of the island magnetization re-orientation. We start from a coverage of $0.5$ ML where the hydrogen is present only on the surface in a $2$H-$(2\times2)$ superstructure, and progressively insert H atoms in octahedral sites within the island. The insertion of $1$ or $2$ H atoms (corresponding respectively to a coverage of $0.75$ and $1$ ML) can be carried out at different locations among the $4$ available sites in the $2\times2$ cell. We retain the configurations displaying in-plane anisotropy as in the experiment, and energetically most stable. We find that up to $1$ ML coverage, it is more favorable to load the Co/Cu interface. The two corresponding values of MAE have been reported on Fig.~\ref{F4}c as green circles. Increasing the concentration above $1$ ML, involves inserting H at octahedral sites in both Co/Co and Co/Cu interfaces, leading to a limited number of configurations with in-plane magnetization. For concentrations $>1.5$ ML (with $0.5$ ML located at the surface), the MCA energy becomes positive for all possible configurations. 

%The insertion of $1$ or $2$ H atoms (corresponding respectively to a coverage of $0.75$ and $1$ ML) can be carried out at different locations among the $4$ available sites in the $2\times2$ cell. 

%Ultimately, however, only one possibility for each coverage leads to an increase of the MCA and \textit{a fortiori} of the total magnetic anisotropy. 

Having identified potential configurations, it becomes feasible to estimate the hydrogen loading in the Co bilayer by computing the corresponding spin density maps and comparing them to the exchange maps. Figure~\ref{F5}a shows a computed spin density map with an extra 0.25 ML of hydrogen at the octahedral sites of the Co/Cu interface. There is no significant difference compared to the experimental map. However, increasing the hydrogen to $0.50$ ML at the Co/Cu interface (Fig.~\ref{F5}b) and further hydrogen loading (Fig.~\ref{F5}c) result in spin density patterns that diverge from the experimental observations (Fig.~\ref{F3}b). Based on these findings, if hydrogen is present in the island, the total coverage must be between $0.5$ ML and $0.75$ ML, with $0.5$ ML located on the surface.\\

\noindent\textbf{CONCLUSIONS}\\
In summary, our study demonstrates that exposing a nanoscale magnet to sufficient hydrogen can induce a 90$^{\circ}$ rotation of its magnetization. Although our DFT computations are conducted on infinite Co layers, they successfully reproduce the interplay between hydrogen loading and magnetism observed in finite-size Co magnets, typically few tens of nm wide in our measurements. This rotation is primarily driven by hydrogen adsorption on the magnet's surface and is further reinforced by the presence of hydrogen at the Co/Cu interface. While hydrogen adsorption also occurs at step edges in the experiment, our findings suggest that this effect can be neglected to first order.\\

\noindent\textbf{MATERIALS AND METHODS}\\
\noindent\textbf{Experimental details}\\
We employed a customized Omicron ultrahigh vacuum STM with a pressure maintained below $10^{-10}$ mbar. Inelastic tunneling spectroscopy was performed at 2.4 K, while all other measurements were carried out at $4.4$ K. The Cu(111) substrate was cleaned through multiple cycles of Ar$^+$ sputtering and annealing at a temperature of $520^{\circ}$C. The W tip used in this investigation was first cleaned by Ar$^+$ sputtering. Cobalt islands were grown on Cu(111) by evaporating Co onto the surface at room temperature. Cobalt was evaporated from a rod that had undergone thorough out-gassing, with an evaporation rate of 0.3 ML min$^{-1}$. After a deposition of $1$ ML, the cobalt sample was transferred to the pre-cooled STM. Nickelocene was then deposited by exposing the sample, maintained below 100 K, to a molecular flux of $2.5\times10^{-2}$ ML min$^{-1}$ for a few seconds. 

For the Nc-tip preparation, we first indented the tungsten tip into the copper surface, resulting in the creation of a copper-covered, mono-atomically sharp tip apex. Subsequently, we positioned the tip above an Nc molecule adsorbed on either a cobalt or copper step edge. To attach Nc to the tip, the tunneling parameters were set at $-1$ mV and $50$ pA, and the the tip was carefully approached towards the molecule by a minimum of 200 pm. Notably, molecular attachment was also possible with Nc molecules adsorbed on the Co islands, employing tunneling parameters of $50$ mV and $50$ pA, and approaching the tip by $350$ pm. Details concerning Nc-tip characterization can be found elsewhere \cite{Fetida2024}. The $dI/dV$ spectra were recorded using a lock-in amplifier operating at a frequency of $6.2$ kHz and a modulation of $5$ mV rms. The $d^2I/dV^2$ were recorded using a lower modulation of $500$ $\mu$V rms.

Following the deposition of nickelocene, we initiated the hydrogenation of the sample. Hydrogen molecules constitute the predominant residual gas in the UHV chamber. In our UHV environment, where the pressure is $<5\times10^{-11}$ mbar, hydrogen contaminates the Co islands over time. Achieving the desired hydrogen superstructure on the islands requires however an ``efficient'' hydrogenation. This consists in increasing the temperature of the cryostat to more than $17$ K for less than 1 hour in order to prompt the release of H$_2$ from the gold-plated copper walls of the STM cryostat \cite{Natterer2013b}.\\

\noindent\textbf{Computational details}\\
DFT calculations were performed using the VASP code \cite{kresse_efficiency_1996}. The PBE \cite{perdew_generalized_1996} form of GGA was used as an exchange and correlation functional. Core electrons were treated following the PAW method \cite{kresse_ultrasoft_1999}. Two supercell structures were used, both representing epitaxially extended Co bilayers on a $(111)$ oriented Cu substrate. The first one is a $2\times2-(111)$ supercell including $2$ layers of Co ($4$ Co atoms per layers) and $9$ layers of underlying Cu ($4$ Cu atoms per layer). It enables to achieve surface hydrogen coverages between $0.25$ and $1$ ML by the addition of $1$ to $4$ hydrogen atoms in hollow sites. Possible insertion in the most stable inner octahedral sites was also considered with $4$ octahedral sites available at the CoCo and CoCu interfaces, \textit{i.e.} $8$ in total or $12$ including the hollow surface adsorption sites. The second geometry used is the simple $1\times1-(111)$ supercell including $2$ layers of Co ($1$ Co atom per layer) and $9$ layers of underlying Cu ($1$ Cu atom per layer). In this case, the addition of one hydrogen atom at the surface or at the CoCo and CoCu interfaces in interstitial position, corresponds to full hydrogen layers. A vacuum layer of 10~{\AA} was always added in the direction perpendicular to the surface in order to minimize interactions introduced by periodical conditions. In any case an energy cutoff of $500$ eV is used. Then for the $2\times2$ cell a $12\times12\times1$ $k$-point sampling was applied while a larger $30\times30\times1$ $k$-point sampling was used for the smaller $1\times1$-cell.  Considering pristine Co and full adsorbed or inserted H layers, we checked that using one cell geometry or the other, with its associated k-point set, led to the same magnetocrystalline anisotropy energies with an error of less than $20$ $\mu$eV per atom.

Calculations are performed into two main steps following the magnetic force theorem \cite{steiner_calculation_2016}, one involving spin-polarized calculations in the collinear scheme and the other including spin-orbit coupling. In the collinear case, the supercell is relaxed along the $z$ direction while the lateral lattice constant is fixed to the one of Cu ($a=3.635$ {\AA}, obtained after optimization of the copper substrate alone) according to an epitaxial growth. The positions of all atoms except for those in the two bottom layers were relaxed (along z direction) until all forces were smaller than $0.01$ eV/{\AA} and the total energy converged within an accuracy of $1\cdot10^{-7}$ eV. At this stage, the output charge densities are used to deduce the spin-densities, whose maps are plotted with the VESTA software \cite{momma_vesta_2011}, and STM images are simulated following Tersoff-Hamann theory \cite{tersoff_theory_1985} using the STMpw code \cite{lorente_stmpw_2019}. The same charge densities are also used in order to perform calculations including the spin-orbit interaction as implemented in VASP \cite{steiner_calculation_2016}.

The magnetocrystalline anisotropy energy (MCA) was determined by rotating the spins according to different crystallographic directions. In our case, the spin-orbit coupling was taken into account non-self-consistently for the spin orientations corresponding respectively to in-plane and out-of-plane magnetizations. Site- and orbital-resolved energies are provided within the spin-orbit calculations in VASP. More precisely, we get $E_{soc}$ on each ion which represents the accumulated energy contribution inside the augmentation sphere that is centered at each ion position. In order to determine the total magnetic anisotropy, we add to the MCA the so-called shape anisotropy, which results from magnetostatic dipole-dipole interactions and therefore depends on the geometry of the system under study. In the case of the infinite bilayers of the present work, this contribution $E_{dd}$ was evaluated numerically using the following summation up to an in-plane cut-off radius of 150 {\AA} \cite{Hammerling2002}:

\begin{equation}
    E_{dd}=\frac{\mu_0}{8\pi} \sum_{i \neq j} \frac{1}{|\textbf{r}_{ij}|^3} \left[\textbf{m}_i\cdot\textbf{m}_j-3 \frac{(\textbf{r}_{ij}\cdot\textbf{m}_i)\cdot(\textbf{r}_{ij}\cdot\textbf{m}_j)}{|\textbf{r}_{ij}|^2}\right]
    \nonumber
\end{equation}
The total MAE, between out-of-plane ($\perp$) and in-plane magnetization ($\parallel$), is therefore given by the difference of energies:

\begin{equation}
\begin{aligned}
    \text{MAE}  &= \text{MCA}+\Delta E_{dd} \\
                &= (E_{tot,\perp}^{\text{DFT}}-E_{tot,\parallel}^{\text{DFT}})+(E_{dd,\perp}-E_{dd,\parallel})
\end{aligned}  
\nonumber
\end{equation}
Using this latter equation leads to positive (negative) values of MAE for in-plane (out-of-plane) orientation.
\\

%%%%Acknowledgements%%%%
\noindent\textbf{ACKNOWLEDGEMENTS}\\
\noindent The Strasbourg authors acknowledge support from the EU’s Horizon 2020 research and innovation programme under the Marie Skłodowska-Curie grant 847471, from the International Center for Frontier Research in Chemistry (Strasbourg), from project ANR-23-CE09-0036 funded by the ANR, and from the High Performance Computing Center of the University of Strasbourg. Part of the computing resources were funded by the Equipex Equip@Meso project (Programme Investissements d'Avenir) and the CPER Alsacalcul/Big Data. R.R. and N.L. thank financial support from projects RTI2018-097895-B-C44, PID2021-127917NB-I00 funded by MCIN/AEI/10.13039/501100011033, from project QUAN-000021-01 funded by the Gipuzkoa Provincial Council and from project IT-1527-22 funded by the Basque Government. Funded by the European Union. Views and opinions expressed are however those of the author(s) only and do not necessarily reflect those of the European Union. Neither the European Union nor the granting authority can be held responsible for them.

%%%%SUPP INFO%%%%
%\begin{suppinfo}
%Information is provided concerning tip/sample preparation and DFT calculations. Additional experimental and computational data are %presented that support the main conclusions of the article (Figures S1-S7). 
%\end{suppinfo}

%%%%BIBLIOGRAPHY%%%%
%\bibliography{references}% Produces the bibliography via BibTeX.


\end{document}